\documentclass[12pt,twoside]{article}
\usepackage{correl14}

\def\TheTitle{Dynamical Vertex Approximation}
\def\TheHeading{Dynamical Vertex Approximation}
\def\TheAuthor{Karsten Held}
\def\TheInstitute{Institute for Solid State Physics\\ Vienna University of Technology}
\def\TheAddress{1040 Vienna, Austria}
\def\TheChapter{5}                       

\begin{document}
\index{dynamical vertex approximation (D$\rm \Gamma$A)}
\MakeTitle     

                        \section{Introduction}

The theoretical description and understanding of strongly correlated systems
is particularly challenging since  perturbation theory in terms of
the Coulomb interaction is no longer possible
and standard mean field theory does not work.
Also bandstructure calculations in the local density approximation
(LDA) \cite{Jones} which had been so successful for the calculations of
many materials do not work properly as electronic correlations are only
rudimentarily taken into account. 
A big step forward in this respect is  {dynamical mean field theory} \index{dynamical mean field theory (DMFT)} (DMFT)
 \cite{DMFT1,DMFT2,DMFT3,PhysToday}
which is a mean field theory in the spatial coordinates, but fully accounts
for the local correlations  in time (quantum fluctuations). In comparison, standard Hartree-Fock is mean-field in space  {\it and} time.
DMFT is non-perturbative since all Feynman diagrams are taken into account,
albeit only their local contribution for the self energy. If one is dealing with well localized $d$- or $f$-orbitals, these local
DMFT correlations often provide  the major part of the electronic 
correlations, which not only give rise to  mass renormalizations  \cite{DMFT2,DMFT3}, metal-insulator transitions \cite{DMFT2,DMFT3} and magnetic ordering \cite{Jarrell92a,DVZPHYS}, but also to unexpected
new physics such as kinks \cite{kinks,poorman} or the filling of the 
Mott-Hubbard gap  with increasing temperature \cite{Mo02}. More aspects of
DMFT are discussed in other contributions to this J\"ulich Autumn School on DMFT at 25.

The question we would like to address here is: Can we do (or do we need to do) better than DMFT?
Indeed, going beyond DMFT is necessary since many of the 
most fascinating and least understood
physical phenomena such as quantum criticality and superconductivity
originate from {\em non-local} correlations\index{non-local correlations} -- which are by construction beyond DMFT.
And we can:  A first step  to include non-local
correlations beyond DMFT have been cluster approaches such as the
dynamical cluster approximation (DCA) \index{dynamical cluster approximation (DCA)} \cite{Maier04,LichtensteinDCA,Koch} and cluster DMFT  \index{cluster dynamical mean field theory (CDMFT)} (CDMFT) \cite{clusterDMFT,LichtensteinDCA,Koch}. Here, instead of considering
a single site embedded in a mean-field (as in DMFT) one considers
a cluster of sites in a mean-field medium. Numerical limitations however 
restrict
the DCA and CDMFT calculations to about 100 sites.
This allows for studying short-range correlations, particularly for 
two-dimensional lattices, but severely restricts the approach for three
 dimensions, for multi-orbitals, and regarding long-range correlations.

Because of that, in recent years  {\em diagrammatic extensions} of 
DMFT were at the focus of the methodological development.
An early such extension was the  $1/d$ ($d$: dimension) approach \cite{Schiller}; also the combination of the non-local spin fermion self energy with  
the local quantum fluctuations of DMFT has been proposed \cite{Kuchinskii05a}.
Most actively pursued are nowadays diagrammatic approaches based
on the local two-particle  vertex.\index{vertex}
The first such approach has been the  dynamical vertex approximation 
(D$\rm \Gamma$A) \cite{DGA1}, followed by the  dual fermion approach \index{dual fermion (DF) approach} \cite{DualFermion}, the one-particle irreducible approach (1PI) \cite{1PI},
and DMFT to functional renormalization group (DMF$^2$RG) \index{dynamical mean field theory to functional renormalization group  (DMF$^2$RG) approach} \index{functional renormalization group (fRG)} \cite{DMF2RG}.

The very idea of these approaches is to extend the DMFT concept of
taking all (local)  Feynman diagrams for the one-particle
irreducible vertex (i.e., the self energy), to the next, i.e., two-particle 
level. In these approaches, one calculates the local two-particle vertex;  and from this non-local correlations beyond DMFT are obtained diagrammatically. Indeed, we understand 
most (if not all) physical phenomena either on 
the one-particle level (e.g.\ the quasiparticle renormalization and the Mott-Hubbard transition) or on the two-particle level [e.g.\ (para)magnons and (quantum)
critical fluctuations]. Non-local correlations and associated 
physics on this two-particle
are included in these diagrammatic extensions of DMFT, which
however still include the local DMFT one-particle physics 
such as the formation of Hubbard bands and the metal-insulator transition
or, more precisely, a renormalized version thereof.
The concept of all these approaches is similar, but they differ in which two-particle vertex is taken and  which diagrams are constructed, see Table \ref{Table:1}. Depending on the approach, Feynman diagrams are constructed from
 full Green function lines $G(\nu, {\mathbf k})$ or  from the difference 
between  $G(\nu, {\mathbf k})$ and the local Green function  $G_{\rm loc}(\nu)$ [$\nu$:  (Matsubara) frequency; ${\mathbf k}$: wave vector]. The DF, 1PI and DMF$^2$RG approach are also based on a
 generating functional integral.

\begin{table}[tb]
\begin{center}
\begin{tabular}{l | l | l}
Method  & Local two-particle vertex & Feynman diagrams\\
\hline
DF \cite{DualFermion} & one-particle reducible vertex, here$^*$  $F_{\rm loc}$ & 2nd order, ladder\\
&&  parquet\\

\hline
1PI \cite{1PI} & one-particle irreducible vertex $F_{\rm loc}$ & ladder \\
\hline
DMF$^2$RG  \cite{DMF2RG} & one-particle irreducible vertex $F_{\rm loc}$ & RG flow \\
\hline
ladder D$\rm \Gamma$A  \cite{DGA1} & two-particle irreducible vertex in channel $r$ $\Gamma_{r \rm loc}$ & ladder\\
\hline
full D$\rm \Gamma$A  \cite{DGA1}  & two-particle fully irreducible vertex  $\Lambda_{\rm loc}$ & parquet
\end{tabular}
\end{center}
 \caption{Summary of the different diagrammatic extensions of DMFT based on the two-particle vertex. All methods are based on the local  part of the two-particle vertex named in the table; for a definition of the different vertex functions, see Section \protect \ref{Sec:reducible}.\newline
$^*$ Note that at the two-particle level every two-particle vertex is one-particle irreducible; three or higher order vertices can be however  one-particle reducible which has consequences for the  diagrammatics if truncated at the two-particle level, see \protect \cite{1PI}.
\label{Table:1}}
\end{table}

In these lecture notes, we will concentrate on D$\rm \Gamma$A in the following.
Section \ref{Sec:reducible} recapitulates the concept of reducible and irreducible diagrams, the parquet and Bethe Salpter equation. On this basis,
we introduce
in Section \ref{Sec:DGA} the   D$\rm \Gamma$A approach. 
In Section~\ref{Sec:Results} we have chosen two exemplary 
highlights that demonstrate  what can be calculated by D$\rm \Gamma$A and related approaches. These are: the calculation of the critical exponents for the three dimensional
Hubbard model 
(Section~\ref{Sec:Critexp}) and the effect of long-range correlations on the Mott-Hubbard transition for the two dimensional Hubbard model (Section~\ref{Sec:Mott}), at zero temperature antiferromagnetic fluctuations always open a gap at any interaction $U>0$.

\section{Feynman diagrammatics}\index{Feynman diagrams}
\subsection{Parquet equations}
\label{Sec:reducible}
The very idea of D$\rm \Gamma$A is a resummation of Feynman diagrams, not in 
 orders of the interaction $U$ as in perturbation theory but in terms of the locality of diagrams.  In this sense, DMFT
is the first, one-particle level since it approximates the one-particle 
fully irreducible vertex, i.e., the self energy $\Sigma$, to be local, see Table~\ref{Fig:DGAoverview}.
 
\begin{table}[tb]
\begin{center}
\begin{tabular}{| l | l |}
\hline
$n=1$ & DMFT: local self energy \\
$n=2$ & D$\rm \Gamma$A: local fully irreducible two-particle vertex\\
& \hfill  $\Rightarrow$ non-local self energy/correlations \\
$\cdots$ & \\
$n\rightarrow \infty$ & exact solution\\
\hline
\end{tabular}
\end{center}
 \caption{D$\Gamma$A generalizes the DMFT concept of the local self energy (i.e., the local fully irreducible one-particle vertex) to the fully irreducible $n$-particle vertex.  It is hence a resummation of Feynman diagrams in terms of their locality. On the right hand side the different levels of approximation are indicated. \label{Fig:DGAoverview}}
\end{table}

\begin{figure}[tb]
\centering \includegraphics[width=0.65\textwidth]{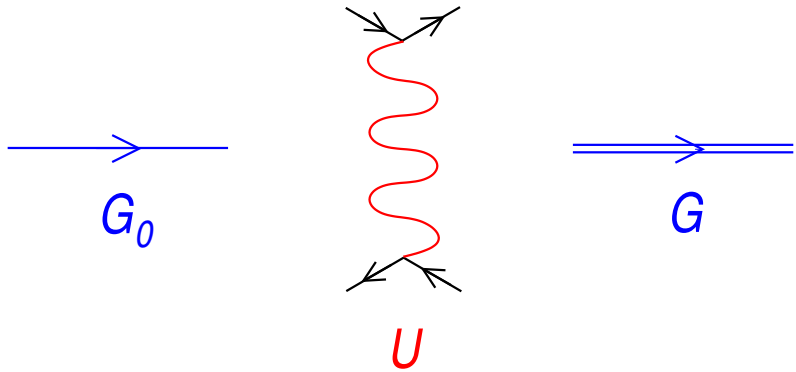}
 \caption{Basic objects of Feynman diagrams: from the non-interacting Green function $G_0$ (left) and
the bare interaction $U$ (middle) we construct all topologically distinct diagrams for calculating the  interacting Green function $G$ (right) \label{Fig:DGAF1}.}
\end{figure}

For a better understanding of reducibility and irreducibility as well as of D$\rm \Gamma$A later on, let us
recall that in quantum field theory we  calculate the interacting
Green function $G$
by drawing all topologically distinct Feynman diagrams that consist of $n$ interactions $U$ and that are connected
by  non-interacting Green function lines $G_0$, keeping one incoming and one outgoing $G_0$ line, see Fig.~\ref{Fig:DGAF1}.
Each  $G_0$ line  contributes a factor $G_0(\nu,{\mathbf k}) =1/(\nu + \mu -\epsilon_{\mathbf k})$ [where $\nu$ denotes the  (Matsubara) frequency, $\mu$ the chemical potential and $\epsilon_{\mathbf k}$ the energy-momentum dispersion relation of the non-interacting problem]
and each interaction (wiggled line) contributes a factor $U$ \footnote{For a ${\mathbf k}$-dependent, i.e. non-local interaction  the factor $U({\mathbf k},{\mathbf k'},{\mathbf k''},{\mathbf k'''}={\mathbf k}+{\mathbf k'}-{\mathbf k''})$ would be ${\mathbf k}$-dependent. }. Here and in the following, we assume a one band model for the sake of simplicity. For an introduction to Feynman diagrams, more details  and  how to evaluate Feynman diagrams including the proper prefactor, we refer the reader to textbooks of quantum field theory such as
 \cite{Abrikosov}; a more detailed presentation including  D$\rm \Gamma$A can also be found in \cite{RohringerPhD}.

\subsubsection*{Dyson equation and self energy}
\begin{figure}[tb]

 \centering \includegraphics[width=\textwidth]{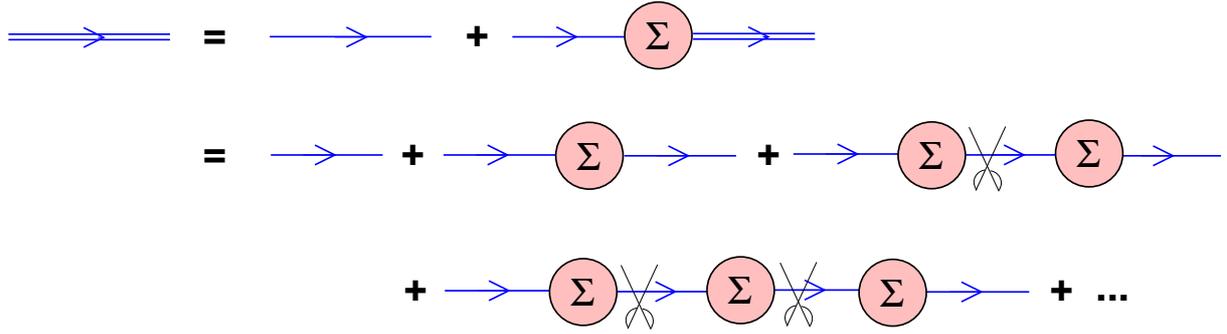}
 \caption{Dyson equation connecting the Green function and self energy. 
The pair of scissors indicates that these diagrams are one-particle reducible (i.e., cutting one $G_0$ line separates the Feynman diagram into two parts) \label{Fig:DGAF2}}
\end{figure}

\begin{figure}[tb]
 \centering \includegraphics[width=0.75\textwidth]{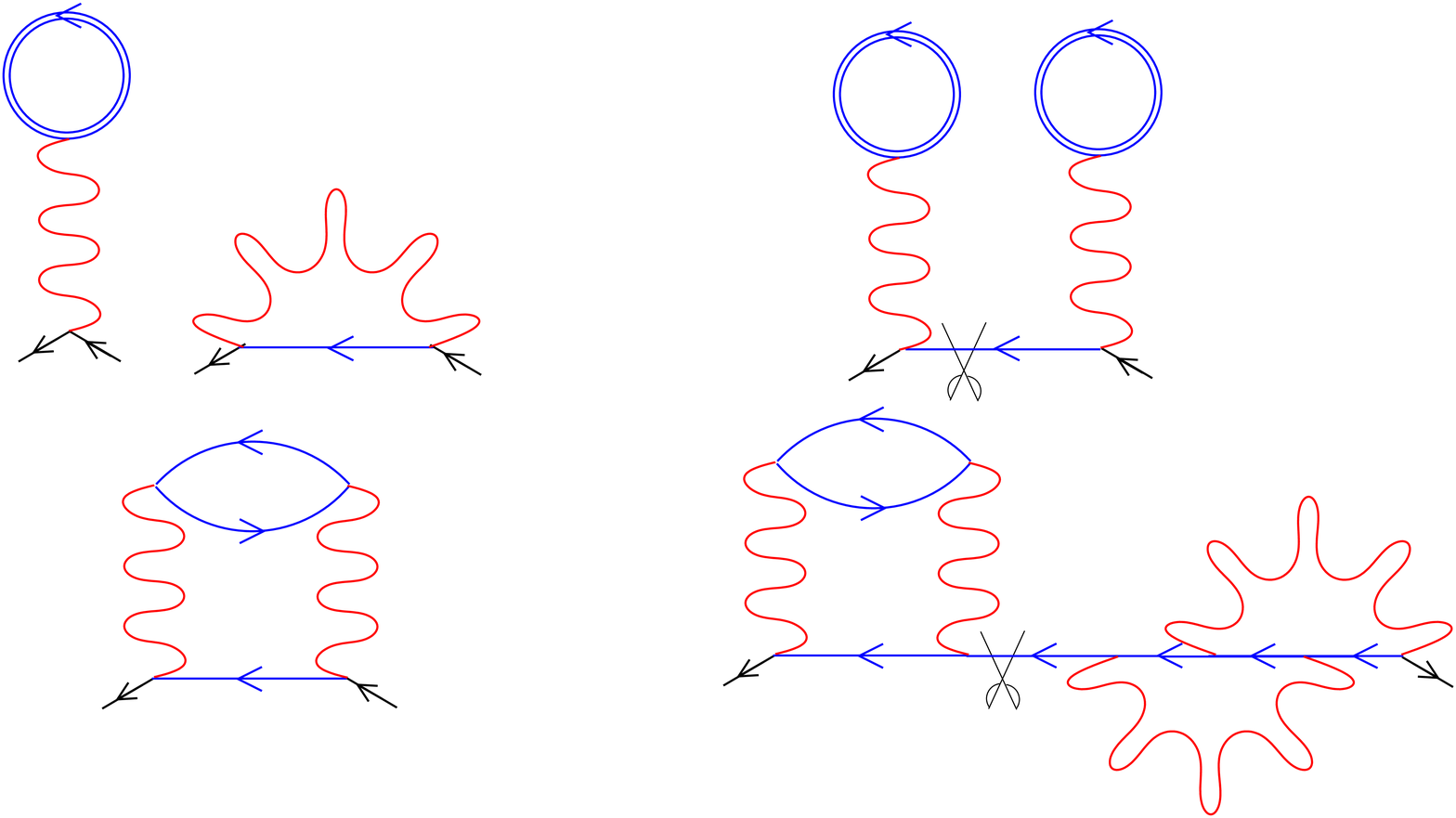}
 \caption{Left: some examples of (one-particle irreducible) self energy diagrams. Right: Diagrams that do no contribute to the self energy since they are one-particle reducible (cutting the line indicated by the pair of scissors separates the diagram into two pieces). \label{Fig:DGAF3}}
\end{figure}

Instead of focusing on the Green function, we can consider a more
compact object, the self energy $\Sigma$,\index{self energy} which is related to $G$ through the Dyson equation, see Fig.~\ref{Fig:DGAF2}.\index{Dyson equation} The Dyson equation can be resolved for the interacting Green function:
\begin{equation}
G(\nu,{\mathbf k})=[1/G_0(\nu,{\mathbf k})-\Sigma(\nu,{\mathbf k})]^{-1} .
\label{Eq:Dyson}
\end{equation}
Since the geometric series of the the Dyson equation generates a series
of Feynman diagrams, we can only include a reduced subset of Feynman diagrams when evaluating the self energy.
One obvious point is that the two outer ``legs'' (incoming and outgoing $G_0$ lines)  are explicitly added when going from the self energy to the Green function, see Fig.~\ref{Fig:DGAF2}.
Hence we have to ``amputate'' (omit) these  outer ``legs'' for self energy diagrams.
More importantly, the self energy can only include one-particle irreducible
diagrams.  
Here, {\em one}-particle (ir)reducible \index{one-particle irreducible} means that by cutting  {\em one} Green function line one can(not) separate the diagram into two parts.
This is since, otherwise, we would generate Feynman diagrams twice:
Any one-particle reducible diagram can be constructed
from two (or more) irreducible building blocks  connected by one (or more)
 single $G_0$ lines. This is exactly what the Dyson equation does, see
 Fig.~\ref{Fig:DGAF2}. For example, in the last line, we have
three irreducible self energy blocks connected by two  single $G_0$ lines.
This shows that, by construction, the self energy has to include 
all  one-particle irreducible Feynman diagrams and no
one-particle reducible diagram. Since the self energy has
{\em one} (amputated) incoming leg, it is a {\em one}
particle vertex. It is also {\em one}-particle irreducible as explained above. Hence, the self energy
is the one-particle irreducible one-particle vertex.
In Fig.\ref{Fig:DGAF3} we show some diagrams that are part of the self energy,
i.e., cutting one-line does not separate the diagram into two parts, and some diagrams  that are not.

\subsubsection*{Two-particle irreducibility}
\begin{figure}[tb]
 \centering \includegraphics[width=0.9\textwidth]{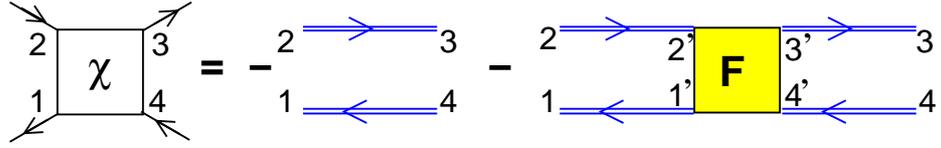}
 \caption{The susceptibility $\chi$ consists of two unconnected Green function $G$ lines (aka ``bubble'') and vertex $F$ corrections. \label{Fig:DGAF4}}
\end{figure}

Let us now turn to the next, the two-particle level.
As illustrated in Fig.~\ref{Fig:DGAF4}, the Feynman diagrams for the susceptibility $\chi$ (or similarly for the 
two particle Green function) \cite{Abrikosov} consist of (i) an  unconnected part (two $G$ lines, as in the non-interacting case)   and (ii) all connected Feynman diagrams (coined vertex corrections). Mathematically this yields ($\beta$: inverse temperature; $\sigma$: spin):
\begin{eqnarray}
\chi_{\sigma \sigma'}(\nu \nu' \omega; {\mathbf k}, {\mathbf k'},  {\mathbf q}) &\!\!=\!\!&  -\beta G(\nu, {\mathbf k})  G(\nu+\omega, {\mathbf k+\mathbf q}) \delta_{\nu \nu'}  \delta({\mathbf k}-{\mathbf k'})\delta_{\sigma\sigma'} 
 \\ \nonumber && + G(\nu, {\mathbf k})  G(\nu\!+\!\omega, {\mathbf k\!+\!\mathbf q}) F_{\sigma \sigma'}(\nu \nu' \omega; {\mathbf k}, {\mathbf k'},  {\mathbf q})  G(\nu', {\mathbf k'})  G(\nu'\!+\!\omega, {\mathbf k'\!+\!\mathbf q}) .\label{Eq:chi}
\end{eqnarray} 
Here $F$ denotes the full, {\em reducible vertex}. 
In the following let us introduce a short-hand notation 
for the sake of simplicity,
where 
$1$ represents a momentum-frequency-spin coordinate $1 \equiv ({\mathbf k},\nu,\sigma)$, $2 \equiv ({\mathbf k + \mathbf q},\nu+\omega,\sigma)$, $3\equiv ({\mathbf k' + \mathbf q},\nu'+\omega,\sigma')$,
$4 \equiv ({\mathbf k'},\nu',\sigma')$. In this notation we have
\begin{equation} 
 \chi(1234) =  - G(14)G(23) -  G(11')G(22')F(1'2'3'4')G(33')G(4'4) \; ,
\end{equation}
 as visualized in Fig.~\ref{Fig:DGAF4}. Since the Green function is diagonal in spin, momentum and frequency, i.e., $G(11')=G(11')\delta_{11'}$, some indices are the same which yields Eq.~(\ref{Eq:chi}).


Let us now again introduce the concept of irreducibility, this time
for the two-particle
vertex. In this case, we  consider {\em two}-particle irreducibility.\footnote{Note, that one-particle irreducibility is somehow trivial since one can show that there are no one-particle irreducible
diagrams for the two-particle vertex (in terms of
the interacting $G$/skeleton diagrams).}
 In analogy to the self energy, we
define the {\em fully irreducible vertex} \index{vertex} $\Lambda$, defined as the set of
all Feynman diagrams  that do not split into two parts 
by cutting two $G$ lines. Let us remark that here and in the following, we construct the Feynman diagrams in terms of $G$ instead of $G_0$. This means that we have to exclude all diagrams 
which contain a structure as generated by the Dyson equation in Fig.~\ref{Fig:DGAF2} since otherwise these diagrams would be counted twice.
This reduced set of diagrams with $G$ instead of $G_0$ is called skeleton diagrams, see \cite{Abrikosov}. 

\begin{figure}[tb]
 \includegraphics[width=\textwidth]{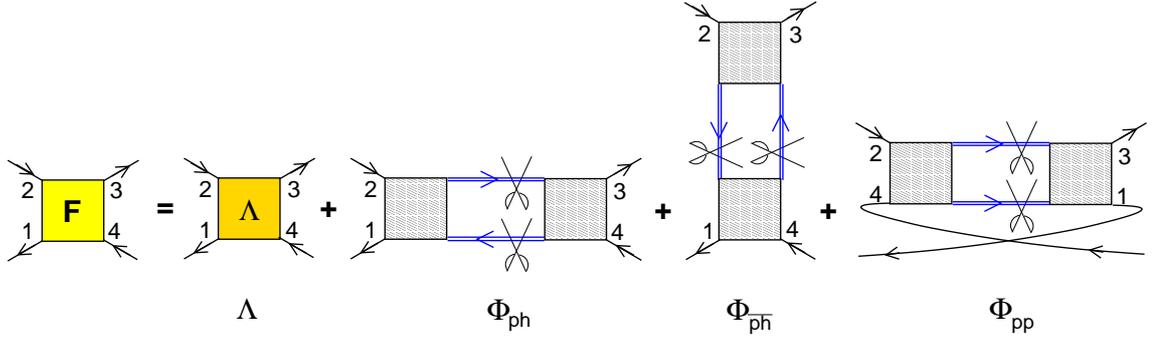}
 \caption{The full (reducible) vertex $F$, consist of the fully irreducible
vertex $\Lambda$ and two-particle reducible diagrams. These can be classified
into three channels depending on which parts  are disconnected by cutting two Green function lines. These are commonly denoted as the
particle-hole reducible channel $\Phi_{ph}$ separating 12 from 34, the 
transversal particle-hole reducible  channel $\Phi_{\overline{ph}}$ separating 14 from 23,
and  the particle-particle reducible  channel $\Phi_{{pp}}$ separating 13 from 24. Each two-particle reducible diagram is reducible in  one (and only one) of these three channels. \label{Fig:DGAF5} The hatched blocks themselves  
can be irreducible, reducible in the same channel$^*$, or reducible in the other two channels (in this last case the full diagram remains however reducible only in the scissors-indicated channel).\newline
$^*$:{Note, this is only possible for  one hatched side, since otherwise the same diagram might be counted twice.}}
\end{figure}

The {\em reducible diagrams} of $F$ can be further classified according
to how the Feynman diagram separates when cutting two internal Green functions. Since $F$ has four (``amputated'')  legs to the outside,
there are actually {\em three} possibilities to split $F$ into two parts by cutting
two $G$ lines. That is, an external leg, say 1, stays connected with one
out of the three remaining external legs but is disconnected from the other two, see 
 Fig.~\ref{Fig:DGAF5} for an illustration. One can show by hands of the diagrammatic topology, that 
each diagram is either fully irreducible or reducible in {\em exactly}
one channel, so that
\begin{equation}
F(12 34)=\Lambda(1234)+\Phi_{ph}(1234)+\Phi_{\overline{ph}}(1234)+\Phi_{{pp}}(1234).\label{Eq:F1}
\end{equation}

\subsubsection*{Bethe-Salpeter equation}
\begin{figure}[tb]
 \centering \includegraphics[width=0.7\textwidth]{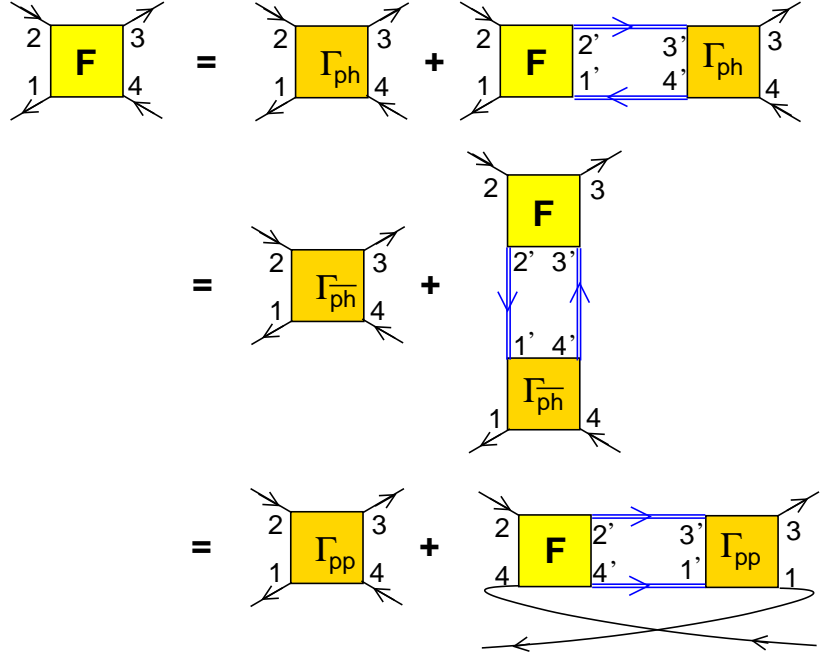}
 \caption{Bethe-Salpeter equation for the particle-hole channel ($ph$, top)
the transversal particle-hole channel($\overline{ph}$, middle) and the
 particle-particle channel ($pp$, bottom).
 \label{Fig:DGAF6}}
\end{figure}

\index{Bethe-Salpeter equation}
We have
 defined the reducible diagrams  $\Phi_{r}$ in channel 
$r \in\{ ph, \overline{ph}, pp\}$ as a subset
 of Feynman diagrams for $F$. The rest, i.e., $F-\Phi_{r}$,
is called  the vertex  $\Gamma_r$ {\em irreducible  in $r$} so that 
\begin{equation}
F(1234)= \Gamma_r(1234) + \Phi_{r}(1234) \; .
\label{Eq:F2}
\end{equation}
In analogy to the Dyson equation, Fig.~\ref{Fig:DGAF2}, 
the reducible vertices $\Phi_r$ in turn can be constructed from 
 $\Gamma_r$. One  $\Gamma_r$ can be connected
 by two $G$s with another  $\Gamma_r$ (which makes this diagram two-particle reducible in the channel $r$). This can be connected again by two $G$s with a third  $\Gamma_r$  etc. (allowing us to cut the two $G$'s 
at two or more different positions).
This gives rise to
a geometric ladder series, the so-called Bethe-Salpeter equation, see
Fig.~\ref{Fig:DGAF6}. Mathematically, these Bethe-Salpeter equations read in the three channels (with Einstein's summation convention):
\begin{eqnarray}
F(1234)&=&\Gamma_{ph}(1234) + F(122'1')  G(3'2') G(1'4') \Gamma_{\rm ph}(4'3'34)\label{Eq:BS1}\\ 
&=&\Gamma_{\overline{ph}}(1234) +  F(2'233') G(2'1') G(3'4')\Gamma_{\overline{ph}}(11'4'4)\label{Eq:BS2}\\
&=&\Gamma_{{pp}}(1234) +  F(4'22'4)  G(2'3') G(1'4')\Gamma_{{pp}}(13'31').
\label{Eq:BS3}
\end{eqnarray}

\subsubsection*{Parquet equations}\index{parquet equations}
Since an irreducible $\Gamma_{r}$ diagram in a channel $r$ is either fully irreducible ($\Lambda$)
or reducible  in one of the two other channels $r'\neq r$ ($\Phi_{r'}$), we can  express $\Gamma_{r}$ as [this also follows directly from 
Eqs.~(\ref{Eq:F1}) and  (\ref{Eq:F2})]:
\begin{equation}
\Gamma_{r}(1234) = \Lambda (1234) + \sum_{r'\neq r} \Phi_{r'} (1234) \; . \label{Eq:Gl}
\end{equation}

\begin{figure}[tb]
 \centering \includegraphics[width=\textwidth]{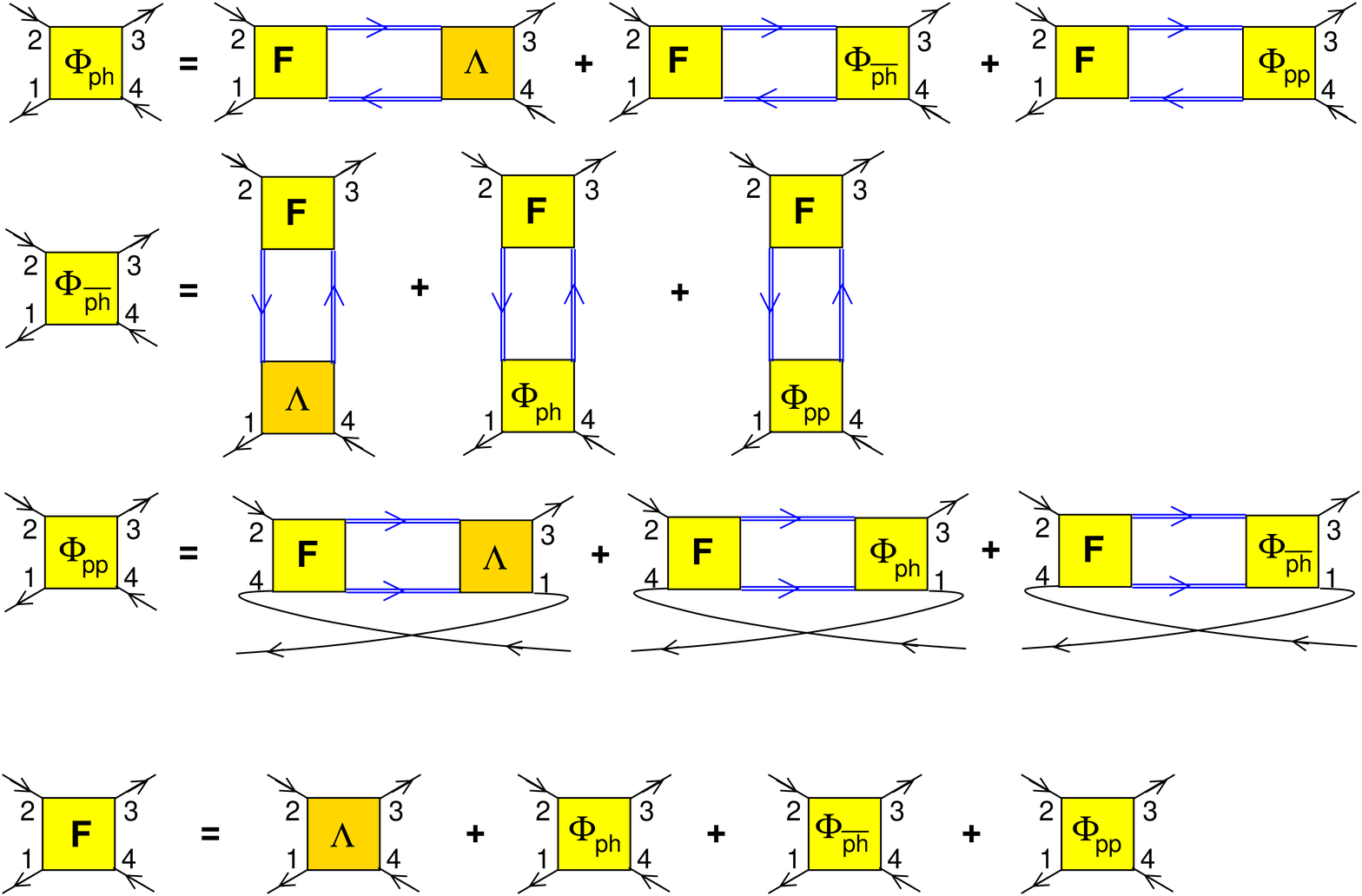}
 \caption{Parquet equations. From the Bethe-Salpeter  equations  in the three channels we obtain
three corresponding equations connecting the reducible vertex $F$, the fully irreducible vertex $\Lambda$ and the reducible vertices $\Phi_r$ in the three  channels $r$ (first three lines). Together with the classification of   $F$ into $\Lambda$ and the three  $\Phi_r$ (last line) we have the   four so-called parquet equations. \label{Fig:DGAF7}}
\end{figure}

We can use this  Eq.~(\ref{Eq:Gl}) to substitute the last
$\Gamma_{r}$'s  in Eq.~(\ref{Eq:BS1}) (or the $\Gamma_{r}$ box in Fig.~\ref{Fig:DGAF6})
 by   $\Lambda$ and $\Phi_r$'s. Bringing the first $\Gamma_{r}$ on the left hand side, then yields:
\begin{eqnarray}
\!\!\!\!\!\!\!\!\Phi_{ph}(1234)&\!\!\!\!=\!\!\!\!&F(1234)-\Gamma_{ph}(1234) \nonumber\\
&\!\!\!\!=\!\!\!\!& F(122'1')  G(3'2') G(1'4')  \Lambda (4'3'34) \!+ \!\!\! \sum_{r'\neq ph} \!\!  F(122'1')  G(3'2') G(1'4')   \Phi_{r'} (4'3'34) 
\label{Eq:parquet2}
\end{eqnarray}
and two corresponding equations for the other two channels.
The corresponding Feynman diagrams are shown in Fig.~\ref{Fig:DGAF7}.
If the fully irreducible vertex $\Lambda$ is known, these three equations
together with  Eq.~(\ref{Eq:F1}) allow us to calculate
the four unknown vertex functions  $\Phi_{r}$ and $F$, see Fig.~\ref{Fig:DGAF7}.
This set of equations is called the {\em parquet equations}\index{parquet equations}.\footnote{Sometimes, also only Eq.~(\ref{Eq:F2}) is called parquet equation and the equations of type Eq.~(\ref{Eq:parquet2}) remain under the name
Bethe-Salpeter equations.}
The solution  can be done numerically by iterating these four {\em parquet equations}. Reflecting how we arrived at the parquet equations, the reader will
realize that the parquet equations are nothing but a classification of  Feynman diagrams into fully irreducible diagrams and diagrams reducible in the three channels $r$.

\begin{figure}[tb]
 \centering \includegraphics[width=.8\textwidth]{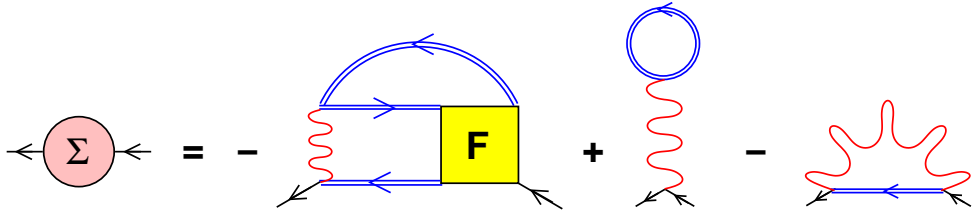}
 \caption{Equation of motion (Schwinger-Dyson equation) for calculating the self energy $\Sigma$ from the
bare interaction $U$, the reducible vertex $F$ and the Green function lines $G$.
The second and third diagram on the left hand side are the Hartree and Fock diagrams, respectively, which are not include in the first term. \label{Fig:DGAF8}}
\end{figure}

The thoughtful reader will have also noticed that the interacting Green function enters  in Eq.~(\ref{Eq:parquet2}).
This $G$ can be calculated by one additional layer of self consistency.
If the reducible vertex $F$ is known, the self energy follows from the Heisenberg
Eq. of motion (also called Schwinger-Dyson Eq. in this context).
This is illustrated in  Fig.~\ref{Fig:DGAF8} and mathematically reads:
\begin{equation}
\Sigma(14) = - U(12'3'1')  G(1'4') G(23') G(2'3) F(4'234) + U(1234)G(23) -  U(1432)G(23) 
\label{Eq:eqofmotion}
\end{equation} 
That is the numerical solution of the four parquet equations has to be be supplemented by the  Schwinger-Dyson Eq.~(\ref{Eq:eqofmotion}) and 
the Dyson Eq.~(\ref{Eq:Dyson}), so that also $G$ and $\Sigma$ are calculated self-consistently.

Let us also note that these general equations while having a simple structure 
in the $1234$ notion can be further reduced for practical calculations: The Green functions are diagonal  $G(3'2')=G(3'3')\delta_{2' 3'}$, there is a severe restriction in spin, there is SU(2) symmetry and  one can decouple the equations into
charge(spin) channels $\Gamma_{ph \uparrow \uparrow} +(-)\Gamma_{ph \uparrow \downarrow}$. A detailed discussion hereof is beyond the scope of these lecture notes. For more details on the parquet equations see \cite{Bickers},
for a derivation of the equation of motion also see  \cite{HeldGW}.

\subsection{Dynamical vertex approximation (D$\rm \Gamma$A)}
\index{dynamical vertex approximation (D$\rm \Gamma$A)}\index{vertex}
\label{Sec:DGA}

Hitherto, everything has been exact. If we know the exact 
fully irreducible vertex $\Lambda$, we can calculate through the
parquet equations Fig.~\ref{Fig:DGAF7} [Eqs.~(\ref{Eq:F2}) and (\ref{Eq:parquet2})]
the full vertex $F$; from this through the Schwinger-Dyson equation of motion (\ref{Eq:eqofmotion}) the self energy $\Sigma$ and through the Dyson Eq.~(\ref{Eq:Dyson}) the Green function $G$. With a  new $G$ we can (at fixed) $\Lambda$ recalculate $F$ etc. until convergence.
Likewise if we know the exact irreducible vertex $\Gamma_r$ in one channel $r$, we can calculate $F$ through the corresponding Bethe-Salpeter Eq. (\ref{Eq:BS1}-\ref{Eq:BS3}) and from this (self consistently) $\Sigma$ and $G$.

\begin{figure}[tb]
 \centering \includegraphics[width=\textwidth]{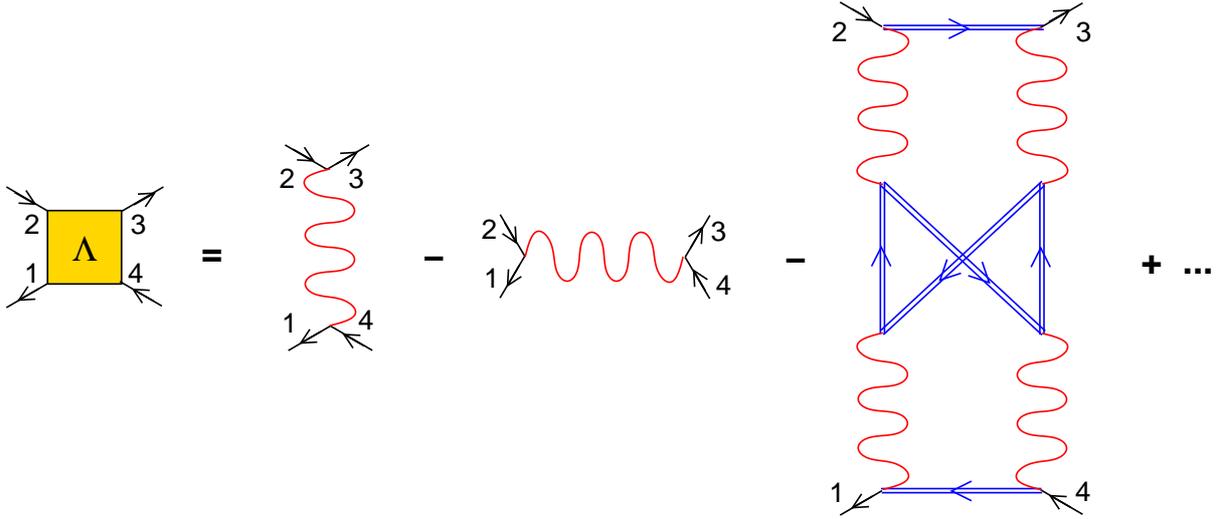}
 \caption{Lowest order Feynman diagrams for the fully irreducible vertex $\Lambda$.
 \label{Fig:DGAF9}}
\end{figure}

But  $\Lambda$ (or $\Gamma_r$) still consists of an infinite set of
Feynman diagrams which we usually do not know.
 Since the parquet (or Bethe-Salpeter) equations
generate many additional diagrams, there are however much less (albeit still infinitely many)
diagrams for  $\Lambda$ (or $\Gamma_r$) than for $F$. In the case of $\Lambda$,  the bare interaction $U$ is included
but the next term is already a diagram of fourth (!) order in $U$ (the so-called envelope diagram), see Fig.~\ref{Fig:DGAF9}. There are no two-particle fully irreducible diagrams of second or third order in $U$.

One approach is hence to approximate  $\Lambda$ by the bare interaction $U$ only, i.e., the first two terms  in  Fig.~\ref{Fig:DGAF9}. This is the so-called parquet {\em approximation} \cite{Bickers}.
For strongly correlated electrons this will be not enough. A deficiency is, for example, that 
 the parquet approximation does not yield Hubbard bands.  

In  D$\rm \Gamma$A, we hence take instead
all Feynman diagrams for $\Lambda$ but restrict ourselves to their local contribution, $\Lambda_{\rm loc}$. This approach is non-perturbative in the local interaction $U$.
It is putting the DMFT concept of locality to the next, i.e., to the two-particle
level. We can extend this concept to the $n$-particle fully irreducible vertex, so that by increasing $n$ systematically more and more Feynman diagrammatic contributions are generated; and for $n\rightarrow \infty$ the exact solution is recovered, see Table~\ref{Fig:DGAoverview} above.

In practice, one has to truncate this scheme at some $n$, hitherto at the two-particle vertex level ($n=2$). The local  fully irreducible two-particle vertex $\Lambda$ can be calculated by solving an Anderson impurity model which has the same local $U$ and  the same Green function $G$. This is because such an Anderson impurity model yields exactly the same (local) Feynman diagrams  $\Lambda_{\rm loc}$.
It is important to note that the locality for  $\Lambda$ is much better fulfilled
than that for $\Sigma$. Even in two dimensions, $\Lambda$
is essentially ${\mathbf k}$-independent, i.e., local. This has been demonstrated by numerical calculations for the two-band Hubbard model, see \cite{Maier06}.
In contrast, for the same set of parameters $\Sigma$ is strongly  ${\mathbf k}$-dependent,  i.e., non-local.
Also $\Gamma_r$ and $F$ are much less local than $\Lambda$,  see \cite{Maier06}.
There might be parameter regions in two dimensions or one-dimensional models where  also  $\Lambda$ exhibits a sizable non-local contributions. One should keep in mind that D$\rm \Gamma$A at the $n=2$ level is still an approximation.  This approximation includes however not only DMFT but on top of that also non-local correlations on all length scales
so that important physical phenomena can be described; and even in two dimensions  substituting  $\Lambda$ by its local contribution  $\Lambda_{\rm loc}$ is a good approximation, better than replacing $\Sigma$ by its local contribution  $\Sigma_{\rm loc}$ as in DMFT.

When solving the Anderson impurity model numerically, one does not
obtain   $\Lambda_{\rm loc}$ directly but first the local susceptibilities $\chi_{\rm loc}$ (or the two-particle Green function). Going from here to  $\Lambda_{\rm loc}$ is possible as follows:
From  $\chi_{\rm loc}$ and  $G_{\rm loc}$, we obtain the local reducible vertex $F_{\rm loc}$ [via  the local version of Eq.~(\ref{Eq:chi})], from this in turn we get   $\Gamma_{ r \rm loc}$ [via inverting the local version of the Bethe-Salpeter Eqs.~(\ref{Eq:BS1}-\ref{Eq:BS3})],  $\Phi_{r\rm loc}$ [via the local version of Eq.~(\ref{Eq:F2})], and finally 
$\Lambda_{\rm loc}$ [via the local version of Eq.~(\ref{Eq:F1})].

\begin{figure}[tb]
 \centering \includegraphics[width=0.5\textwidth]{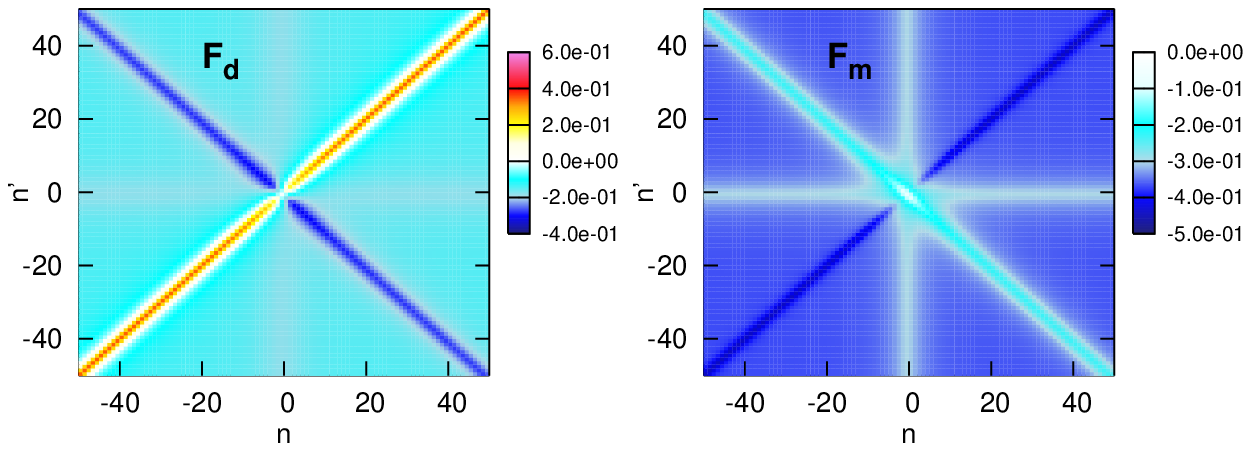}
\vspace{.2cm}

 \centering \includegraphics[width=\textwidth]{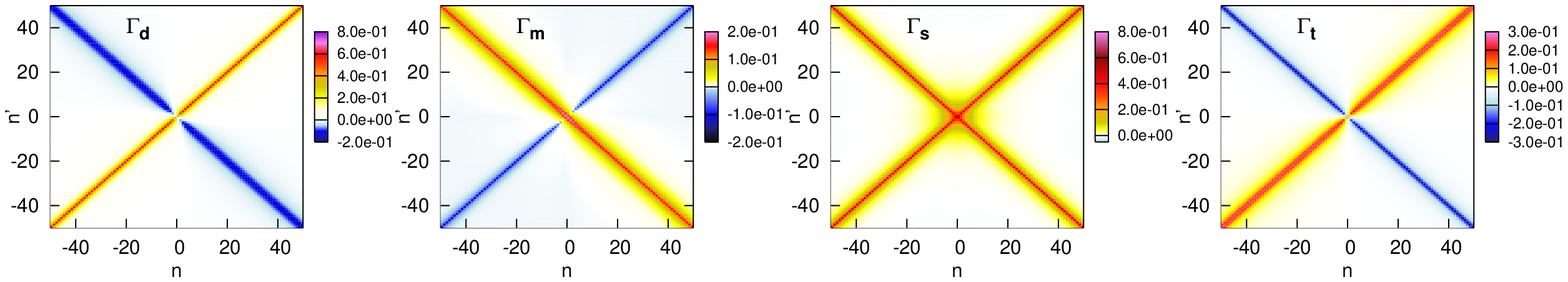}
\vspace{.1cm}

\centering \includegraphics[width=0.5\textwidth]{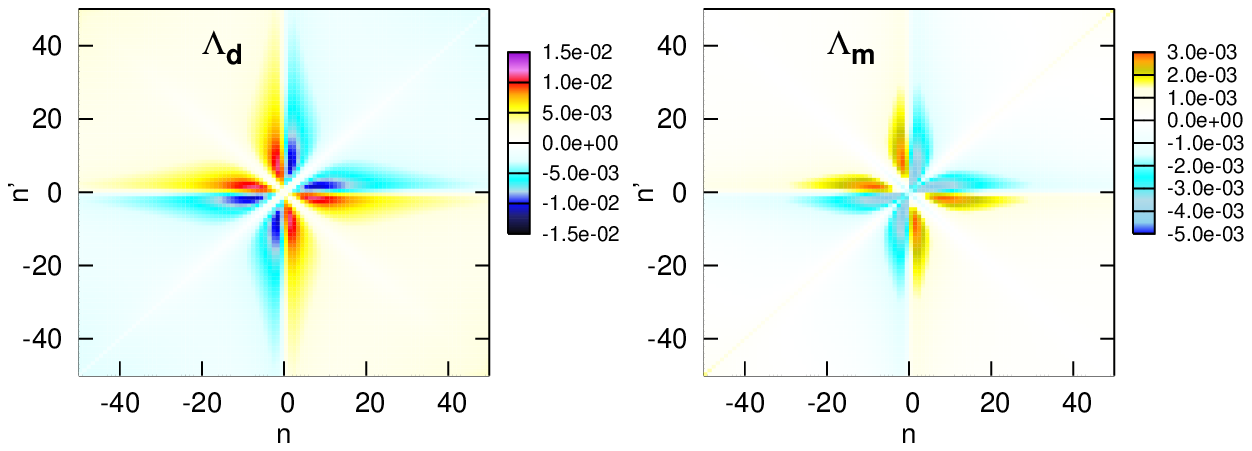}

 \caption{Top: Local full vertex $F$  in the magnetic (m) and charge or density (d) channel, i.e., $F_{\uparrow\uparrow}\pm F_{\uparrow\downarrow}$, as a function
of the two incoming Fermionic frequencies $\nu=(2n+1)\pi T$ and $\nu'=(2n'+1)\pi T$ at transferred Bosonic frequency $\omega=0$ for the three dimensional Hubbard model at  $U=0.5$,
$T=1/26$ (in units of nearest neighbor hopping  $2 \sqrt{6} t\equiv 1$).
 Middle: Corresponding particle-hole irreducible vertex $\Gamma_{ph}$ (two left panels)  and
particle-particle vertex in the singlet (s) and triplet (t) spin combination. The transversal particle-hole channel follows from  $\Gamma_{ph}$ by (crossing) symmetry.
Bottom:  Fully irreducible vertex $\Lambda_{m(d)}$ for the two spin combinations. For all figures, the bare interaction $U$ has been subtracted from the vertices 
(reproduced from \cite{Rohringer12}).
 \label{Fig:DGAF10}\label{Fig:LambdaHM}}
\end{figure}

Fig.~\ref{Fig:LambdaHM} shows the reducible vertex $F$, the irreducible vertex in the particle-hole channel $\Gamma_{ph}$ and the  fully irreducible vertex $\Lambda$ for the three-dimensional Hubbard model on a simple cubic  lattice.
For more details on the calculation of $\Lambda_{\rm loc}$, see \cite{Rohringer12} and  \cite{RohringerPhD}.
Also note that  $\Gamma_{ r \rm loc}$ diverges at an interaction strength $U$ below the Mott-Hubbard metal-insulator transition, which signals  
 the breakdown of perturbation theory \cite{Schaefer13}.

\subsubsection{Self consistency}
After calculating   $\Lambda_{\rm loc}$, we can calculate the full vertex 
$F$ 
through the parquet equations,  Fig.~\ref{Fig:DGAF7}; and through the 
Schwinger-Dyson equation, Fig.~\ref{Fig:DGAF8}, the
non-local self energy $\Sigma$ and Green function $G$,
as discussed in Section \ref{Sec:reducible}.  
Hitherto, all
D$\rm \Gamma$A calculations have stopped at this point. That is, $F$ and $G$ are determined self-consistently but $\Lambda_{\rm loc}$ is not recalculated.

\begin{figure}[tb]
 \centering \includegraphics[width=0.7\textwidth]{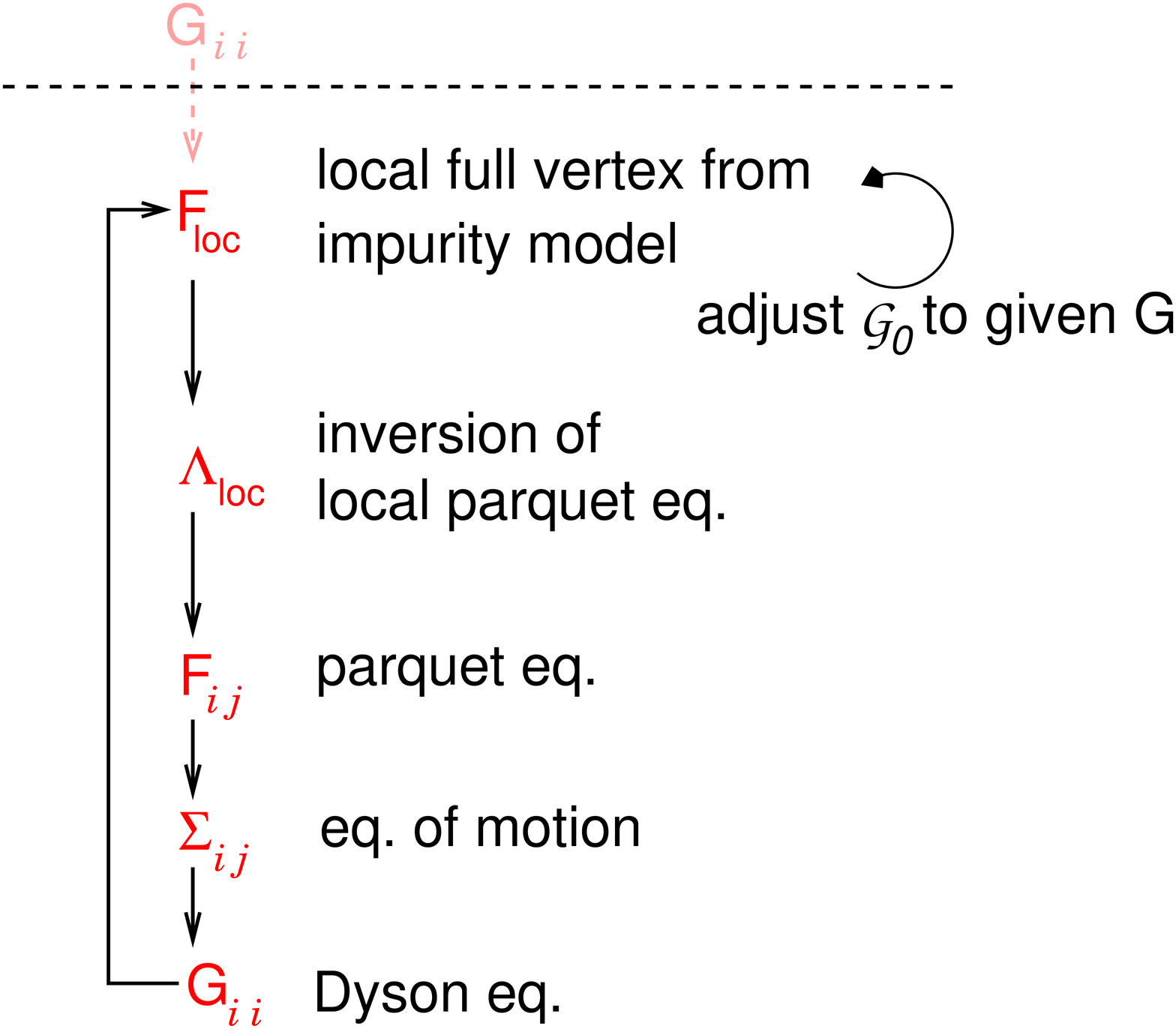}
 \caption{Flow diagram of the D$\rm \Gamma$A approach. Starting from a test Green function $G$ (e.g.,  that of DMFT),
the local susceptibility and local full vertex $F_{\rm loc}$ is calculated by solving an Anderson impurity model.
To this end, the non-local Green function ${\cal G}_0$ of an Anderson impurity  
is adjusted by Eq.~(\ref{Eq:adjustG}) until this  ${\cal G}_0$  impurity model has the given interacting $G$.
From $F_{\rm loc}$ in turn, the inversion of the parquet and Bethe-Salpeter equations allow the calculation of the local fully irreducible vertex $\Lambda_{\rm loc}$.  This is the input of  the parquet equations for calculating the 
non-local vertex $F$ and through the equation of motion and the Dyson equation, the  D$\Gamma$A self energy $\Sigma$ and Green function $G$. In a self-consistent calculation a new local vertex is calculated from $G$  etc. until convergence.
 \label{Fig:DGAFlow}}
\end{figure}

However, in principle, one can self-consistency iterate the 
approach. From the new $G$ we can calculate a new $G_{\rm loc}$.
From this and $U$ we obtain a new vertex etc. until convergence,
see Fig.~\ref{Fig:DGAFlow}. This self-consistency cycle is similar as that of the DMFT but now includes self-consistency on the two-particle ($\Lambda$) level.

Please note that the Anderson impurity model has now to be calculated with
the  interacting $G_{\rm loc}$ from D$\rm \Gamma$A which is different from
the  $G_{\rm loc}$ of DMFT. As  numerical approaches solve the Anderson impurity model 
for a given non-interacting Green function ${\cal G}_0$, we need to adjust this  ${\cal G}_0$ until the  Anderson impurity model's Green function 
 $G$ agrees with the  D$\rm \Gamma$A $G_{\rm loc}$. This is possible
by iterating  ${\cal G}_0$ as follows:
\begin{equation}
 \left[ {\cal G}_0^{\rm new}(\nu)\right]^{-1} =\left[  {\cal G}_0^{\rm old}(\nu)\right]^{-1}+ \left[ G_{\rm loc}(\nu)\right]^{-1} - \left[ G^{\rm old}(\nu)\right]^{-1} \; ,
\label{Eq:adjustG}
\end{equation}
until convergence. Here,  ${\cal G}_0^{\rm old}$ and $G^{\rm old}$ denote the
non-interacting and interacting Green function of the Anderson impurity model from the previous iteration. This  ${\cal G}_0$-adjustment 
is indicated in Fig.~\ref{Fig:DGAFlow} by the secondary cycle.

\begin{figure}[tb]
\centering \includegraphics[width=0.7\textwidth]{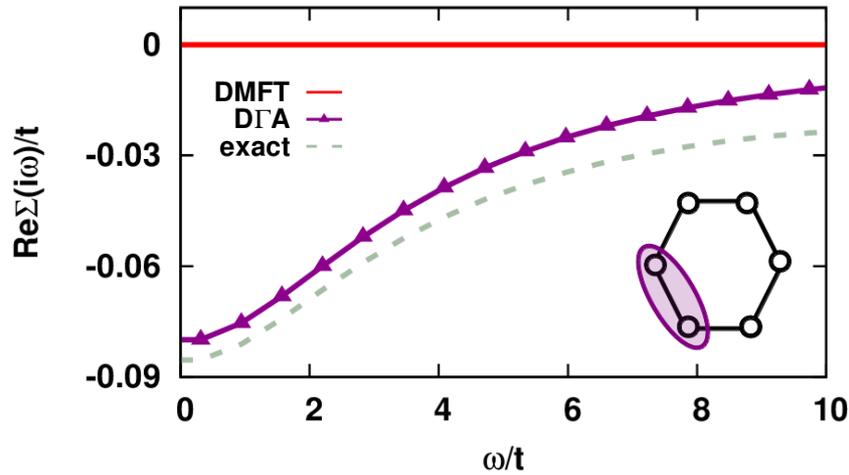}
 \caption{Self-energy $\Sigma_{i i+1}$  between neighboring sites on a six site Hubbard ring (see inset), comparing  DMFT (zero non-local $\Sigma_{i i+1}$), D$\rm \Gamma$A and the exact solution. Parameters: $U=2t$, $T=0.1t$ with $t$ being the nearest neighbor hopping on the ring (reproduced from \cite{ValliPhD}). 
 \label{Fig:Benzene}}
\end{figure}

Fig.~\ref{Fig:Benzene} compares the  D$\rm \Gamma$A self energies calculated this way, i.e.,  the  D$\rm \Gamma$A full parquet solution, with  DMFT and the exact solution.
The results are for a simple one-dimensional Hubbard model with nearest neighbor hopping, six sites and periodic boundary conditions so that the exact solution
is still possible by an exact diagonalization of the Hamiltonian. This can be considered as a simple model for a benzene molecule.
The   D$\rm \Gamma$A  results have been obtained in a ``one-shot'' calculation
 with the DMFT $G$ as a starting point. The good agreement between 
 D$\rm \Gamma$A and the exact solution shows that D$\rm \Gamma$A can be employed for quantum chemistry calculations of correlations in molecules, at least if there is a gap [HOMO-LUMO gap between the highest occupied molecular orbital (HOMO)
and the lowest unoccupied molecular orbital (LUMO)]. For molecules with degenerate ground states (i.e., a peak in the spectral function at the Fermi level), the
agreement is somewhat less impressive; note that  one dimension is the worst possible case for   D$\rm \Gamma$A.

\begin{figure}[tb]
 \centering \includegraphics[width=0.55\textwidth]{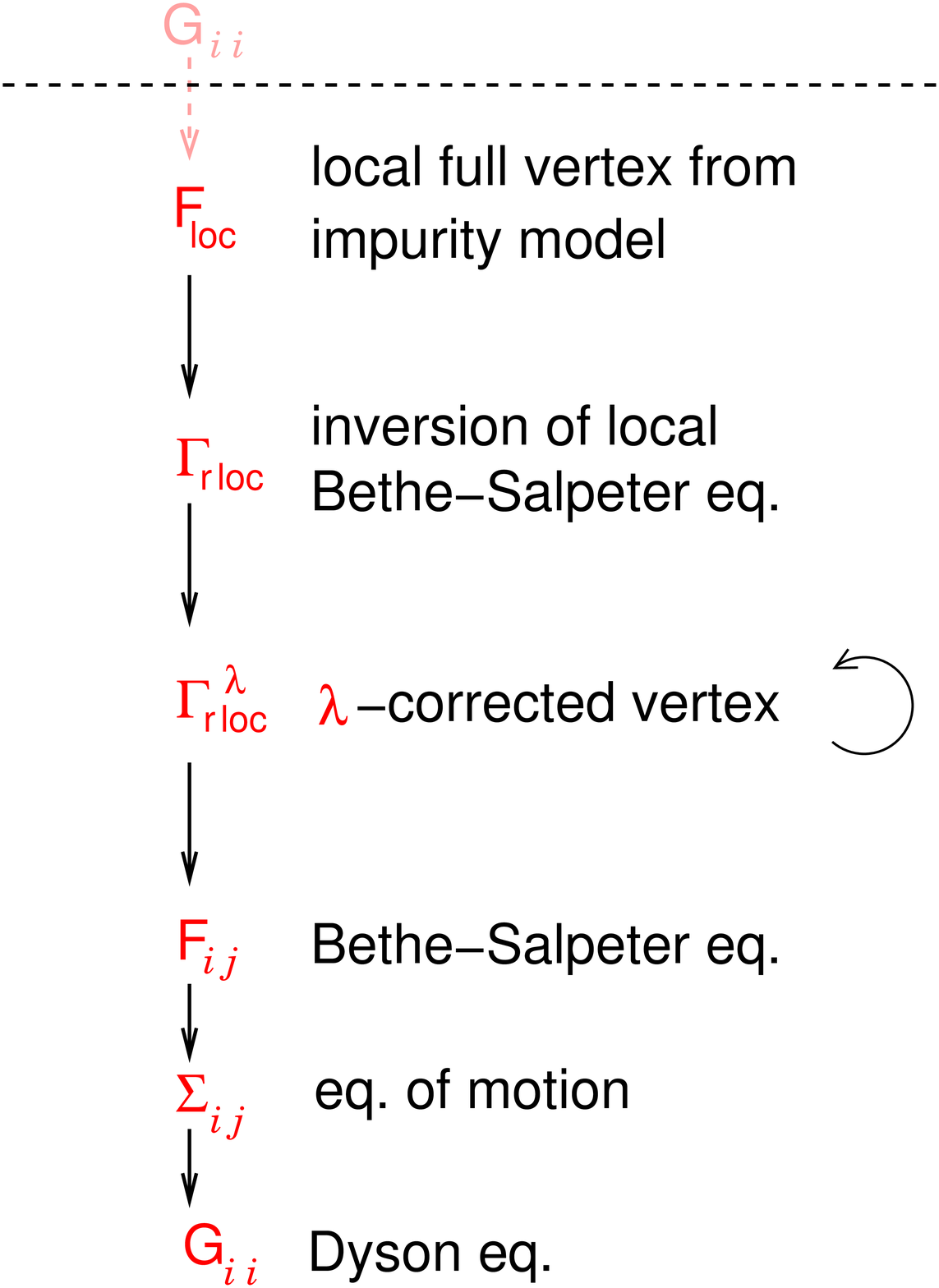}
 \caption{Flow diagram of the D$\rm \Gamma$A ladder approach. Same as Fig.~\ref{Fig:DGAFlow} but solving the Bethe-Salpeter equation(s) with the local irreducible vertex $\Gamma_{r\rm loc}$ in (two) channel(s)
$r$ instead of the parquet equations with the fully irreducible vertex $\Lambda$.
Instead of the self-consistency, it is better to employ a so-called Moriya $\lambda$-correction in this case.
 \label{Fig:DGAFlow2}}
\end{figure}

In Fig.~\ref{Fig:Benzene}, the first parquet D$\rm \Gamma$A results have been shown. However, most  D$\rm \Gamma$A calculations
hitherto employed a simplified scheme based on ladder diagrams, see Fig.~\ref{Fig:DGAFlow2}.
These calculations neglect one of the three channels in the parquet equations:
the particle-particle channel. Both the particle-hole ($ph$) and the transversal  particle-hole channel  ($\overline{ph}$) are taken into account. These
channels decouple for the spin and charge vertex and can hence be
calculated by solving the simpler Bethe-Salpeter equations instead of the full parquet equations. If one neglects non-local contributions in one of the channels, it is  better to restrict oneself
to non-self consistent calculations since part of the neglected diagrams
 cancel with diagrams generated by the self consistency. Instead
one better does a ``one shot'' calculation 
mimicking the self-consistency  by a
 so-called $\lambda$ correction, see \cite{Katanin} for details.
Physically, neglecting the particle-particle channel is justified if non-local
fluctuations of particle-particle type are not relevant. Whether this is the case or not depends on the model and parameter range studied. Such
particle-particle  fluctuations are e.g.\ relevant in the vicinity of superconducting instabilities, where they need to be considered. In the vicinity of antiferromagnetic order  on the other hand, the two particle-hole channels are the relevant ones (and only their magnetic spin combination). These   describe antiferromagnetic fluctuations, coined paramagnons in the paramagnetic phase above the  antiferromagnetic transition temperature.
The D$\rm \Gamma$A results in the next section are for the half-filled Hubbard model. Here we are not only away from any superconducting instability,
but at half-filling the interaction also suppresses  particle-particle fluctuations. Hence, for the results presented below using the ladder instead of the full parquet approximation is most reasonable.

\section{Two highlights}
\label{Sec:Results}
\subsection{Critical exponents of the Hubbard model}
\label{Sec:Critexp}
\begin{figure}[tb]
 \centering
 \includegraphics[width=0.7\textwidth]{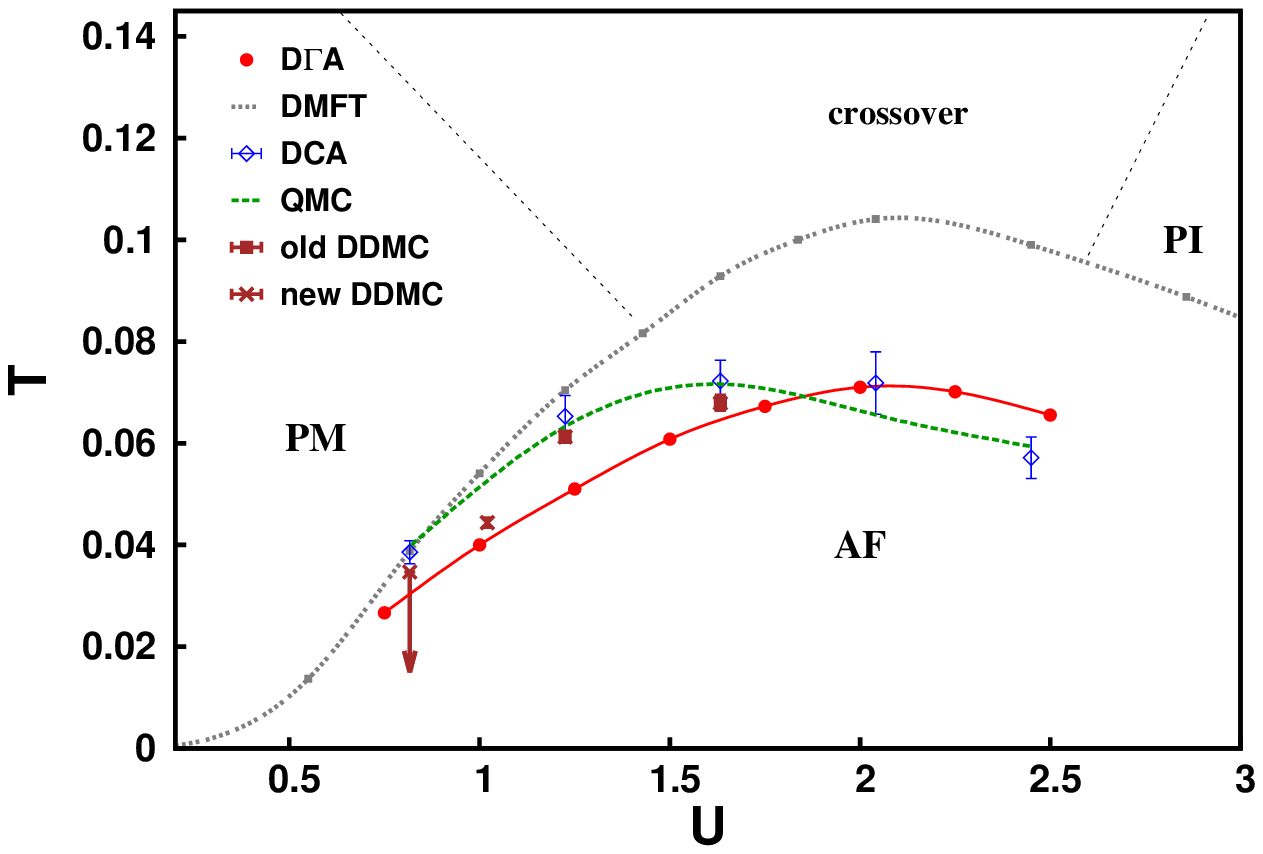}
 \caption{Phase diagram of the Hubbard model \index{Hubbard model} on a cubic lattice with nearest neighbor hopping $ 2 \sqrt{6}t \equiv 1$. The dashed black lines
show the DMFT N\'eel temperature $T_N$
and the DMFT crossover region from a paramagnetic metal (PM) to a paramagnetic insulator (PI). Non-local correlations \index{non-local correlations} reduce $T_N$ with good  agreement between 
D$\Gamma$A \cite{Rohringer11}, DCA \cite{Kent05} \index{dynamical cluster approximation (DCA)}, lattice quantum Monte Carlo (QMC) \cite{Staudt00},
as well as determinantal diagrammatic Monte Carlo (DDMC) before \cite{Gull} and after \cite{TN_new} our D$\Gamma$A results. Note, for the lowest
$U$ value,  Ref.~\cite{TN_new} could only give an upper bound for $T_N$ which according to DDMC could be much smaller as indicated by
the arrow.
 \label{Fig:TN}
}
\end{figure}

The Hubbard model\index{Hubbard model} is the prototypical model for strong electronic correlations.
At half filling it shows an antiferromagnetic ordering at low enough temperatures $T$ --  for all interaction strengths $U>0$ if the lattice has perfect nesting.
Fig.~\ref{Fig:TN} shows the phase diagram of the half-filled three-dimensional Hubbard model (all energies are in units of $D=2 \sqrt{6}  t\equiv 1$,
which has the same standard deviation as a Bethe lattice with half bandwidth $D$). At weak interaction strength, we have a Slater antiferromagnet which can be described by Hartree-Fock theory yielding an exponential increase of the N\'eel temperature with $U$. At strong interactions, we have preformed spins with a Heisenberg interaction $J=4t^2/U$ yielding a Heisenberg antiferromagnet with $T_N\sim J$. In-between these limits,  $T_N$ is maximal. 

All of this can be described by DMFT. However, since DMFT is mean-field w.r.t.\ the spatial dimensions, it overestimates   $T_N$. This can be overcome by including non-local correlations, i.e., spatial (here antiferromagnetic) fluctuations. These reduce $T_N$. In this respect, there is a good agreement between D$\rm \Gamma$A, DCA and lattice QMC, see Fig.~\ref{Fig:TN}. The biggest deviations are observed for the smaller interaction strength. In principle, these differences might originate from the fact that the D$\rm \Gamma$A calculations are not yet self-consistent
and only use the Bethe-Salpeter Eqs.~(\ref{Eq:BS1},\ref{Eq:BS2}) in the two particle-hole
channels instead of the  full parquet equations (\ref{Eq:F2},\ref{Eq:parquet2}). On the other side, however, we observe that long-range correlations are particularly important at weak coupling,
cf.\ Section \ref{Sec:Mott}. Such long-ranged correlations cannot be captured by the cluster extensions
of DMFT or lattice QMC since these are restricted to maximally $\sim 10$ sites in all three directions. More recent and accurate DDMC calculations on larger clusters 
\cite{TN_new} indeed show a smaller $T_N$ and, in contrast to DCA and lattice QMC  better agree with  D$\rm \Gamma$A, see Fig.~\ref{Fig:TN}.

\begin{figure}[t!]
 \centering
 \includegraphics[width=0.7\textwidth]{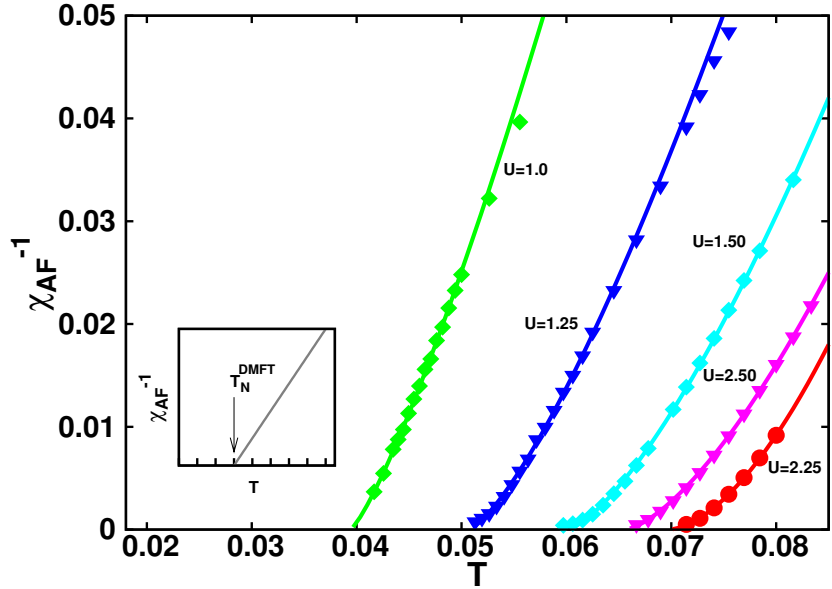}
\caption{Inverse antiferromagnetic  spin susceptibility as a function of $T$ for different interactions $U$ as obtained by  D$\rm \Gamma$A  (lower inset: DMFT), showing in the vicinity of the phase transition a  $\chi_{\rm AF}\sim (T-T_N)^{-2 \nu}$ behavior, with a critical exponent\index{critical exponents}
$\nu \sim 0.7$ in agreement with the three dimensional Heisenberg model \cite{Heisenberg}.
(reproduced from \cite{Rohringer11} \copyright\ by the American Physical Society).
 \label{Fig:CritExp}}
\end{figure}

Even more important are long-range correlations in the immediate vicinity of the phase transition and for calculating the critical exponents. Each finite cluster will eventually show a mean-field exponent. In this respect, we 
could calculate  for the first time the critical exponents of the Hubbard model \cite{Rohringer11}. In Fig.\ \ref{Fig:CritExp} we show the antiferromagnetic
spin susceptibility which shows a
 mean-field-like behavior $\chi_{\rm AF}\sim (T-T_N)^{-1}$ at high temperature
(and in DMFT). In the vicinity of the phase transition however, long-range correlations
become important and yield another critical exponent   $\chi_{\rm AF}\sim (T-T_N)^{-2 \times 0.7}$ which agrees with that of the three dimensional Heisenberg model  \cite{Heisenberg} (as is to be expected from universality).
In contrast for the Falikov-Kimball model, the critical exponents calculated by the related dual Fermion approach \cite{FKM} agree with those of the Ising model.

Except for the immediate vicinity of the phase transition DMFT is
nonetheless yielding a reliable description of the paramagnetic phase.
At least this is true for one-particle quantities such as the self energy
and spectral function, while the susceptibility \cite{Rohringer11} and entropy \cite{Gull} show deviations in a larger $T$-interval above $T_N$.

\subsection{Fate of the false Mott-Hubbard transition in two dimensions}
\label{Sec:Mott}

\begin{figure}[tb]
 \centering
 \includegraphics[width=0.7\textwidth]{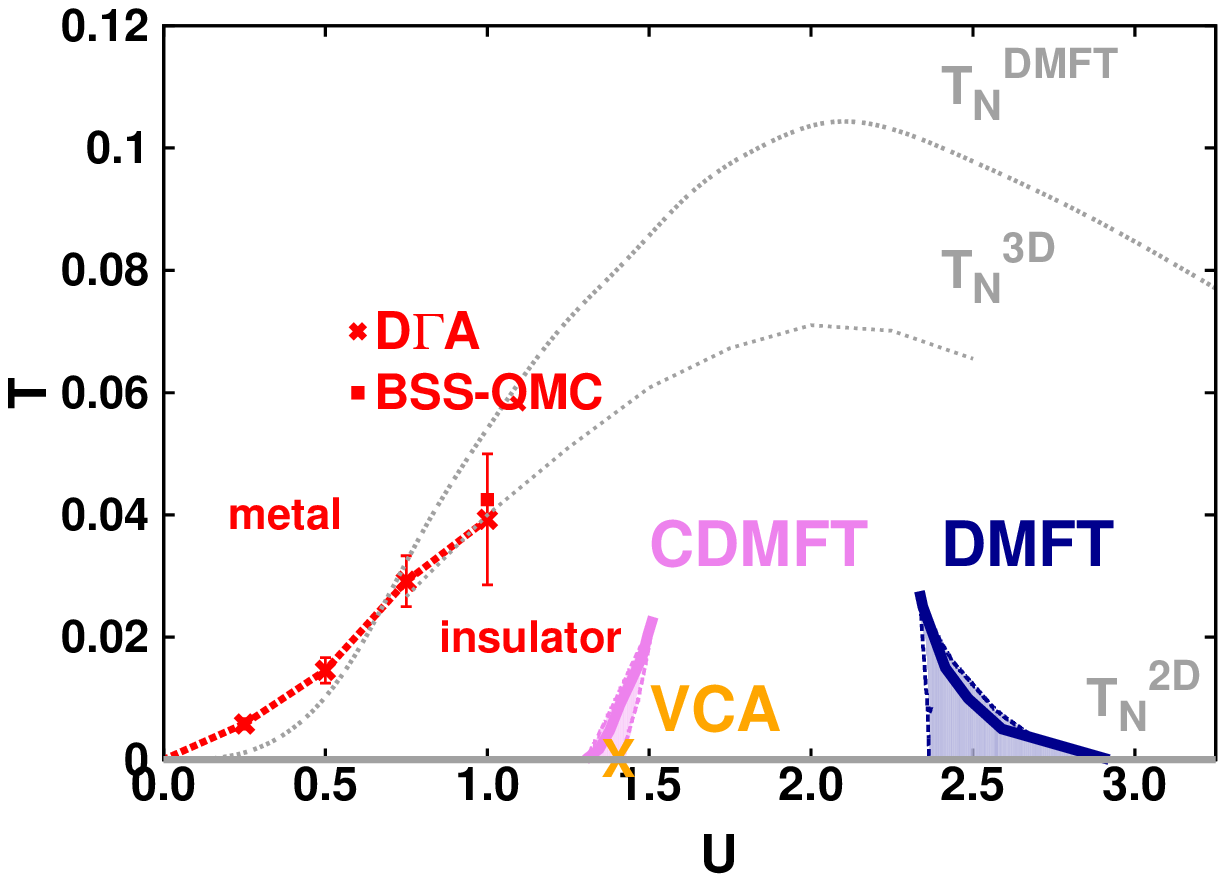}
\caption{Temperature $T$ vs.\ interaction $U$ phase diagram of the two dimensional Hubbard model \index{Hubbard model} on a square lattice with nearest neighbor hopping (all energies are in units of $D=4t\equiv 1$, yielding the same standard deviation as for the 3D phase diagram). 
From local correlations (DMFT \index{dynamical mean field theory (DMFT)} \cite{bluemer_thesis})
via short-range correlations (CDMFT \index{cluster dynamical mean field theory (CDMFT)} \cite{MIT_CDMFT} and
variational cluster approximation (VCA) \index{variational cluster approximation (VCA)} \cite{Schaefer14})
to  long-range correlations (D$\Gamma$A \index{dynamical vertex approximation (D$\rm \Gamma$A)} \cite{Schaefer14}   and BSSQMC \index{quantum Monte Carlo (QMC)} \cite{Schaefer14})
the critical interaction strength for the  metal-insulator transition is reduced to $U_c=0$. The light gray lines denote the DMFT \cite{Kent05}
and D$\rm \Gamma$A \cite{Rohringer11} $T_N$
(reproduced from \cite{Schaefer14}).
 \label{Fig:Mott}}
\end{figure}

As a second example, we recapitulate results for the interplay between antiferromagnetic fluctuations and the Mott-Hubbard transition \index{Mott-Hubbard transition} in two dimensions.
Even though the Hubbard model has been studied for 50 years, astonishingly
little is known exactly. In one dimension it can be solved by the Bethe ansatz, and there is no Mott-Hubbard transition for the half-filled Hubbard model:
For any finite interaction it is insulating. In infinite dimensions, on the other hand DMFT provides for an  exact (numerical) solution. It has been one of the big achievements of DMFT to clarify the nature of the Mott-Hubbard transition which 
is of first order at a finite interaction strength 
\cite{DMFT2,DMFT3}, see 
Fig.~\ref{Fig:Mott}.
 From cluster extensions of DMFT it has been
concluded that the  Mott-Hubbard transition is actually at somewhat smaller $U$ values and the coexistence region where two solutions can be stabilized is smaller, see  Fig.~\ref{Fig:Mott}. However, again these cluster extensions are restricted to {\em short-range} correlations. In particular  at low temperatures, there are strong {\em long-range} antiferromagnetic spin fluctuations
which for example at $U=0.5$ and $T=0.01$ exceed 300 lattice sites \cite{Schaefer14}. The physical origin are antiferromagnetic fluctuations emerging above the
antiferromagnetically ordered phase. In two dimensions this antiferromagnetic phase is restricted to $T=0$ due to the Memin-Wagner theorem, but antiferromagnetic
fluctuations remain strong even beyond the immediate vicinity of the 
 ordered phase (at $T=0$). These long-range antiferromagnetic spin fluctuations
(paramagnons) give rise to pseudogap physics, where first only
part of the Fermi surface becomes gaped but at lower temperatures 
the entire Fermi surface is gaped so that we have an insulating phase
for any $U>0$. Fig.~\ref{Fig:Mott} shows the development from
local correlations (which only yield an insulating phase only relatively large $U$'s) to additional short-range correlations (which reduce the critical $U_c$ for the Mott-Hubbard transition)
to long-range correlations (which reduce $U_c$  to zero).
At large $U$, we have localized spins which can be described by a spin model or the Mott insulating phase of DMFT. At smaller $U$ we also have an insulator
caused by antiferromagnetic spin fluctuations. 
These smoothly
go over into the $T=0$ antiferromagnetic phase, which is
of Slater-type for small $U$. Since the correlation are exceedingly
long-ranged, the nature of the low-temperature gap 
is the same as the Slater antiferromagnetic gap, even though there 
is no true antiferromagnetic order yet.

\section{Conclusion and outlook}
In the last years we have seen the emergence of diagrammatic extensions of DMFT. All these approaches have in common that they calculate a local vertex and
construct  diagrammatically non-local correlations from this vertex.
In regions of the phase diagram where non-local correlations are short-range,
results are similar as for cluster extensions of DMFT. However, the diagrammatic extensions also offer the opportunity to include long-range correlations on an equal footing. 
This allowed us to study critical phenomena and 
to resign  the Mott-Hubbard transition in the two-dimensional
Hubbard model to its fate (there is no Mott-Hubbard transition).

These were just the first steps. Indeed the diagrammatic extensions
offer a new opportunity to address the hard problems of solid state 
physics, from superconductivity and quantum criticality to quantum phenomena
 in nano- and heterostructures. Besides a better
physical understanding by hands of model systems, also realistic materials
calculations are possible -- by {\em Abinitio}D$\rm \Gamma$A\cite{AbinitioDGA}. 
Taking the bare Coulomb interaction and all local vertex corrections as a starting point, {\em Abinitio}D$\rm \Gamma$A includes  DMFT, $GW$\index{GW approach} and non-local correlations beyond within a common underlying framework. 
Both on the model level and for realistic materials calculations,
there is plenty of physics to explore.

{\em Acknowledgments.} I sincerely thank my coworkers S.~Andergassen,  A.~Katanin, W.~Metzner, T.~Sch\"afer, G.~Rohringer, C.~Taranto,  A.~Toschi, and A.~Valli for the fruitful cooperations on diagrammatic extensions of DMFT in the last years, and G.~Rohringer also for reading the manuscript. 
This work has been financially supported in part by  the European Research Council under the European Union's Seventh Framework Programme (FP/2007-2013)/ERC through grant agreement n. 306447 ({\em AbinitioD$\rm \Gamma$A}) and in part by the  Research Unit FOR 1346 of the Deutsche Forschungsgemeinschaft and the Austrian Science Fund (project ID  I597).



\clearchapter


\end{document}